\documentclass[10pt]{book}

\usepackage[T1]{fontenc}
\usepackage[utf8]{inputenc}

\usepackage{mathptmx}

\usepackage{geometry}
\usepackage{url}
\usepackage{bm}

\usepackage{booktabs}

\usepackage{graphicx} 
\usepackage{amssymb,amsfonts,amsmath}

\usepackage{natbib}
\usepackage{microtype}

\usepackage{color}

\usepackage{todonotes}

\usepackage[nolist]{acronym}

\usetikzlibrary{calc}

\input{knitr-preamble.tex}

\newcommand{\R}{\emph{R}}
\newcommand{\Rpackage}[1]{{\texttt{#1}}}
\newcommand{\Rfunction}[1]{{\texttt{#1}}}

\newcommand{\Mq}{\Rpackage{MALDIquant}}
\newcommand{\MqF}{\Rpackage{MALDIquantForeign}}

\begin{acronym}[MALDI-TOF]
  \acro{COW}{Correlation Optimized Warping}
  \acro{DTW}{Dynamic Time Warping}
  \acro{FWHM}{Full Width at Half Maximum}
  \acro{LOWESS}{LOcally WEighted Scatterplot Smoothing}
  \acro{MAD}{Median Absolute Deviation}
  \acro{MALDITOF}[MALDI-TOF]{Matrix-Assisted Laser Desorption/Ionization --
  Time-Of-Flight Mass Spectrometer}
  \acro{MS}{Mass Spectrometry}
  \acro{MSI}{Mass Spectrometry Imaging}
  \acro{mz}[\textit{m/z}]{\emph{mass-to-charge ratio}}
  \acro{PF4}{Platelet Factor 4}
  \acro{PQN}{Probabilistic Quotient Normalization}
  \acro{PTW}{Parametric Time Warping}
  \acro{SNR}{Signal-to-Noise Ratio}
  \acro{SNIP}{Statistics-sensitive Non-linear Iterative Peak-clipping algorithm}
  \acro{TIC}{Total Ion Current}
 \end{acronym}

\begin{document}

\setcounter{chapter}{11}

\noindent{\large \textbf{Chapter \arabic{chapter}}}\\
{\Large \textbf{Mass Spectrometry Analysis Using MALDIquant
\footnote{\underline{Please cite as:} Gibb, S. and Strimmer, K. 2016.
\emph{Mass spectrometry analysis using MALDIquant.}
 Chapter 11 in: Datta, S., and Mertens, B. (eds).  2016.
\emph{Statistical Analysis of Proteomics, Metabolomics, and Lipidomics Data Using Mass Spectrometry.}
Frontiers in Probability and the Statistical Sciences.
Springer, New York.}}}

\vspace{1cm}
\noindent\textbf{Sebastian Gibb\footnote{Anesthesiology and Intensive Care Medicine,
  University Hospital Greifswald, Ferdinand-Sauerbruch-Stra\ss{}e,
  D-17475 Greifswald, Germany.  Email: \url{mail@sebastiangibb.de}}
and Korbinian Strimmer\footnote{Epidemiology and Biostatistics, School of Public Health,
  Imperial College London, Norfolk Place, London W2 1PG, UK. Email: \url{k.strimmer@imperial.ac.uk}}}
\vspace{2mm}

\noindent\emph{Version 20th November 2015}

\vspace{1cm}

\noindent\textbf{Abstract}
\Mq\ and associated R packages provide a versatile and completely free open-source platform for analyzing 2D
mass spectrometry data as generated for instance by MALDI and SELDI instruments.  We first describe the various methods and algorithms available in \Mq.
Subsequently, we illustrate a typical analysis workflow using \Mq\ by investigating an experimental cancer data set,
starting from raw mass spectrometry measurements and ending at multivariate classification.

\section{Introduction}

\ac{MS}, a high-throughput technology commonly used in proteomics, enables the
measurement of the abundance of proteins, metabolites, peptides and amino acids
in biological samples. The study of changes in protein expression across
subgroups of samples and through time provides valuable insights into cellular
mechanisms and offers a means to identify relevant biomarkers, e.g, to
distinguish among tissue types, or for predicting health status.
In practice, however, there still remain many analytic and computational
challenges to be addressed, especially in clinical diagnostics
\citep{Leichtle2013}.
Among these challenges the availability of open and easy-to-extent processing
and analysis software is highly important \citep{Aebersold2003, Lilley2011}.

Here, we present \Mq{} \citep{MALDIquant}, a complete open-source analysis
pipeline for the \R{} platform \citep{RPROJECT}.  In the first half of this chapter
we describe the methodology implemented and available in \Mq.  In the second half 
we illustrate the versatility of \Mq\ by application to an experimental data set,
showing, how raw intensity measurements are preprocessed and how peaks relevant
for a specific outcome can be identified.

Current documentation of specific version of \Mq\ can be found on its homepage
at \url{http://strimmerlab.org/software/maldiquant/}  where we also 
provide instructions for installing
the software.  In addition, we provide a number of example R scripts on the \Mq\ homepage.
Direct download of the \Mq\ software is also possible from the CRAN server at \url{http://cran.r-project.org/package=MALDIquant}.

\section{Methodology Available in \Mq}

\subsection{General Workflow}

The purpose of \Mq\ is to provide a complete workflow to facilitate the complex
 preprocessing tasks needed to convert
raw two-dimensional \ac{MS} data, as generated for example by MALDI or SELDI instruments,
 into a matrix of feature intensities required for
high-level analysis. A typical workflow
is depicted in figure~\ref{fig:workflow}.

Each analysis with \Mq\  consists of all or some of 
the following steps (see also \cite{Norris2007,
Morris2010} for related analysis pipelines):
First, the raw data is imported into the \R{} environment. Subsequently, the
data are smoothed to remove noise and also transformed for variance stabilization.
Next, to remove chemical background noise a baseline correction is
applied.  This is followed by a calibration step to allow comparison of
intensity values across different baseline-corrected spectra.   As a next step
a peak detection algorithm is employed to identify potential features and also to reduce
the dimensionality of the data.  After peaks have been identified a peak alignment
procedure is applied as the \acp{mz} typically differ across different measurements and need 
to be adjusted accordingly.  Finally, after feature binning an intensity matrix is produced
that can be used as starting point for further statistical analysis, for example
for variable selection or classification.

In the following subsections we discuss each of these steps in more detail.

\begin{figure}[t!]
  \centering
  \resizebox{.9\linewidth}{!}{\footnotesize{\begin{tikzpicture}[node distance=3em]
\tikzstyle{line} = [draw, -latex]
\tikzstyle{header}=[text width=15em, text centered, minimum height=2em, font=\bfseries]
\tikzstyle{box}=[rectangle, draw, text width=10em, text centered, minimum height=2em]
\tikzstyle{mqfbox}=[rectangle, draw, text width=10em, text centered, minimum height=2em]
\tikzstyle{mqbox}=[rectangle, draw, text width=11em, text centered, minimum height=2em]
\tikzstyle{sdabox}=[rectangle, draw, text width=10em, text centered, minimum height=2em]
\tikzstyle{frame}=[rectangle, draw, color=blue, loosely dotted, thick]

\node [box] (rd) {Raw Data};
\node [node distance=12.5em, mqfbox, right of=rd] (di) {Data Import};
\node [node distance=3em, header, above of=di] (mqf) {MALDIquantForeign};
\node [node distance=13em, mqbox, right of=di] (vs) {Smoothing and Transformation};
\node [node distance=3em, header, above of=vs] (mq) {MALDIquant};
\node [mqbox, below of=vs] (bc) {Baseline Correction};
\node [mqbox, below of=bc] (cb) {Intensity Calibration};
\node [mqbox, below of=cb] (pd) {Peak Detection};
\node [mqbox, below of=pd] (pa) {Peak Alignment};
\node [mqbox, below of=pa] (pb) {Peak Binning};
\node [mqbox, below of=pb] (fm) {Feature Matrix};
\node [node distance=13em, sdabox, right of=fm] (cl) {Post Processing};
\node [node distance=3em, box, below of=cl] (result) {Result};

\draw [frame] ($(di.north west)+(-0.20, 1)$) rectangle ($(fm.south east) + (-3.6, -0.2)$);
\draw [frame] ($(vs.north west)+(-0.20, 1)$) rectangle ($(fm.south east) + (0.20, -0.2)$);

\path [line] (rd) -- (di);
\path [line] (di) -- (vs);
\path [line] (vs) -- (bc);
\path [line] (bc) -- (cb);
\path [line] (cb) -- (pd);
\path [line] (pd) -- (pa);
\path [line] (pa) -- (pb);
\path [line] (pb) -- (fm);
\path [line] (fm) -- (cl);
\path [line] (cl) -- (result);

\end{tikzpicture}}}
  \caption[MALDIquant workflow]{Preprocessing workflow for \ac{MS} data
  using \MqF{} and \Mq{}.}
  \label{fig:workflow}
\end{figure}

\subsection{Import of Raw Data}

A prerequisite of any analysis is to import the raw data into the \R{} environment.
Unfortunately, nearly every vendor of mass spectrometry machinery has its
own native and often proprietary data format. This complicates the exchange of
experimental data between laboratories, the use of analysis software and the
comparison of results. Fortunately,  there is now much effort to create generic
and open formats, such as \emph{mzXML} \citep{Pedrioli2004} and its successor
\emph{mzML} \citep{Martens2011} or \emph{imzML} \citep{Schramm2012} for \ac{MSI}
data. Nevertheless, the support of these formats is still limited and often conversion is
needed to get the data into a suitable format for subsequent analysis
\citep{Chambers2012}.

Importing of raw data in \Mq\ is performed by its sister R package \MqF{}. 
It offers import routines for numerous native and public data formats. 
In addition to the open XML formats
(\emph{mzXML}, \emph{mzML}) it supports \emph{Ciphergen XML}, \emph{ASCII},
\emph{CSV}, \emph{NetCDF}, and \emph{Bruker Daltonics *flex Series} files. It
can also read \ac{MSI} formats like \emph{imzML} and \emph{ANALYZE 7.5}
\citep{Robb1989}.

A very useful feature of \MqF{} is that it reads and traverses whole directory trees 
containing supported file formats so that simultaneous import of many spectra is 
straightforward.  Furthermore, \MqF{} allows to import data from 
remote resources so the spectral data can be read  over an Internet connection from a website or database.

After importing the raw spectra an important step is quality control.
This includes checking the mass range, the length of each spectra and 
also visual exploration of spectra to find and remove potentially
defective measurements.
\Mq\ provides functions to facilitate this often neglected task. 

\subsection{Intensity Transformation and Smoothing}

The raw data obtained from mass spectrometry experiments are counts of ionized
molecules, with  intensity values approximately following a \emph{Poisson} distribution
\citep{Skoeld2007, Du2008}.  Consequently, the variance depends on the mean,
as  mean and variance are identical for a
\emph{Poisson} distribution. 
However, by applying a square root transformation ($ f(x) = \sqrt{x}$) we can convert the
Poisson distributed data to approximately normal data, with constant variance 
independent of mean, which is an important requirement for many statistical
 tests \citep{Purohit2003}.
In the preprocessing noise models other than the Poisson may be also assumed, which lead to different
variance-stabilizing functions
 such as the logarithmic transformation \citep{Tibshirani2004, Coombes2005}.
These can be easily applied in  \Mq\ as well.

Subsequently, the transformed spectral data is smoothed to 
reduce small and high-frequent variations and noise. For this purpose
 \Mq{} offers the moving average smoother and the
\emph{Savitzky-Golay}-filter \citep{Savitzky1964}. The latter is based on
polynomial regressions in a moving window. In contrast to the moving average, the
 Savitzky-Golay filter preserves the shape of the local maxima. 

Note that  both algorithms require 
the specification of window size, which according to
\citet{Bromba1981} should be chosen to be
smaller than twice the \ac{FWHM} of the peaks.

\subsection{Baseline Correction}

The elevation of the intensity values in a typical \ac{MALDITOF} spectrum is
called \emph{baseline} and is caused by chemical noise such as matrix-effects and
pollution.
It is recommended to remove these background effects to reduce their influence
in quantification of the peak intensities.

In the last few years many algorithms to adjust for the baseline have been developed,
ranging from  simple methods like the subtraction of the absolute minimum
\citep{Gammerman2008} or the moving minimum or median \citep{Liu2010} to 
more elaborate methods such as fitting a LOWESS curve, a spline or an exponential
function against the moving minima respectively median values
\citep{Tibshirani2004, Williams2005, PROcess, Liu2009, He2011, House2011}.
Other authors prefer morphological filters such as \emph{TopHat}
\citep{Sauve2004}, iterative methods as the \ac{SNIP} \citep{Ryan1988} or the
convex hull approach \citep{Liu2003}.

\begin{figure}[t!]
  \includegraphics[width=\textwidth]{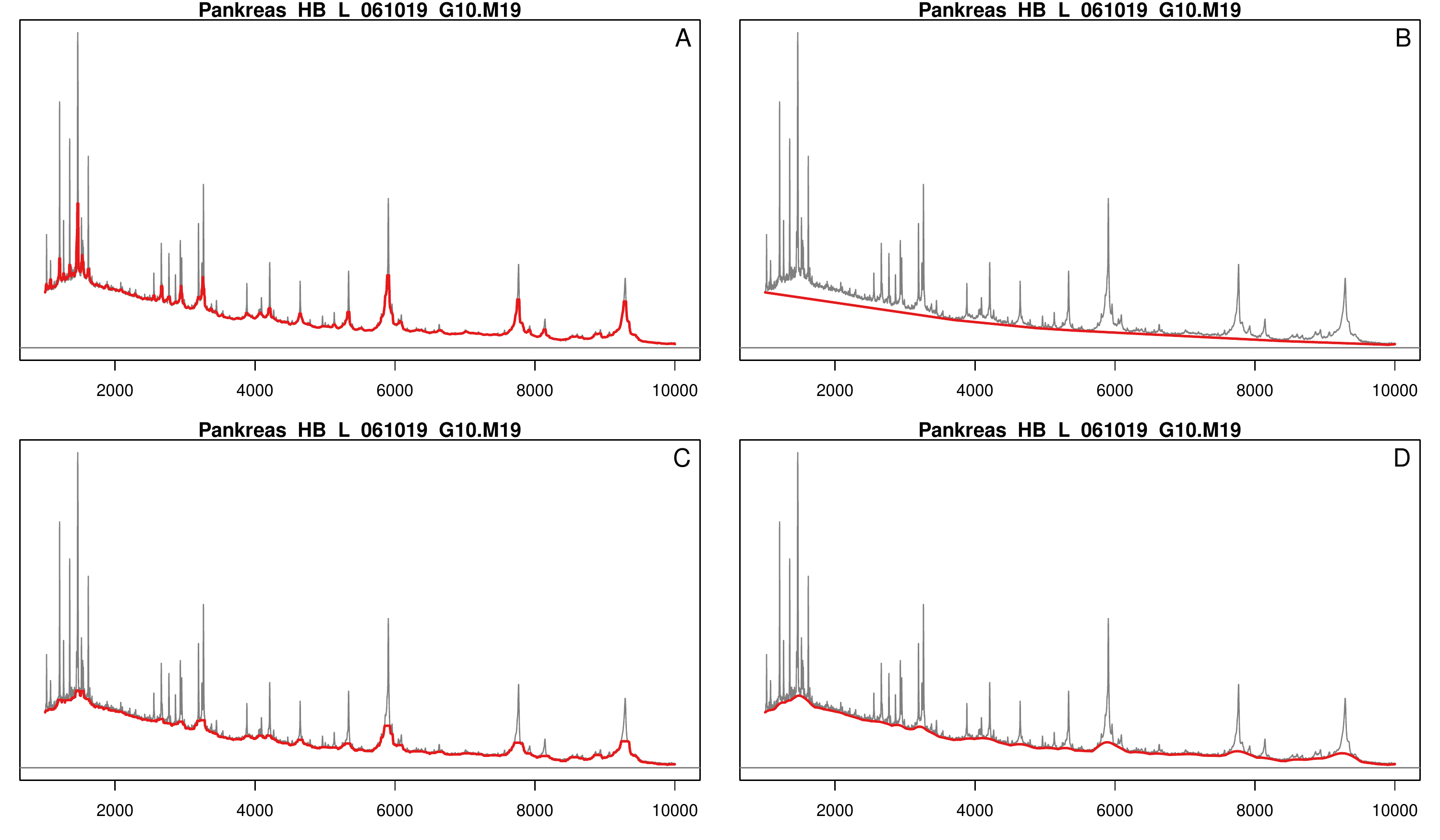}
  \caption[Baseline Estimation]{Estimated baselines for an raw \ac{MALDITOF}
    spectrum from \citet{Fiedler2009}. The following algorithms were applied:
    (A) moving median,
    (B) convex hull,
    (C) \emph{TopHat},
    (D) \ac{SNIP}.}
  \label{fig:be}
\end{figure}

Unfortunately, there is no automatic way to select among the available procedures
to find the baseline correction method that is most suitable for a given spectrum at hand.
Instead, it is recommended to investigate multiple baseline estimations by visual inspection
\citep{Williams2005}.  As shown in Fig.~\ref{fig:be} the algorithms can indeed 
differ substantially.

\Mq{} provides three complex baseline correction algorithms that have been selected
for inclusion in \Mq\ because of their favorable
properties, such as respecting peak form and non-negativity of intensity values:
\begin{enumerate}
\item
The \emph{convex hull} algorithm \citep{Andrew1979}
doesn't need a tuning parameter and is often very effective to find the baseline.
Unfortunately, for concave matrix effects as common in \ac{MALDITOF} spectra
this algorithm cannot be applied --- see  Fig.~\ref{fig:be}B ($\ac{mz} \approx 1500 Da$).
\item
\emph{TopHat} 
\citep{Herk1992, Gil2002} is a morphological filter combining a moving minimum
(erosion filter) followed by a moving maximum (dilation filter). In contrast to
the convex hull approach it has an additional tuning parameter, the window size
of the moving window, that controls the smoothness of the estimated baseline.
The narrower the window the more of the baseline is removed but also of the peak
heights. A wider window will preserve the peak intensities and
produce a smoother baseline but will also cause some local background variation
to remain (Fig.~\ref{fig:be}C).
\item
The default baseline correction algorithm in \Mq{} is \ac{SNIP} \citep{Ryan1988}. 
Essentially, this is a local window-based
algorithm in which a baseline is reconstructed by  replacing the intensities
in a window by the mean of the surrounding points, if the mean is smaller than the
local intensity, with window size decreasing iteratively starting from a specified upper
limit \citep{Morhac2009}.
\end{enumerate}

In addition to the above  \Mq{} also supports the moving-median algorithm (Fig.~\ref{fig:be}A) which is
 commonly used in the literature but may lead to negative intensity values after baseline subtraction.

\subsection{Intensity Calibration}
\label{subsec:ic}

The intensity values in mass spectrometry data represent the relative amount of
analytes, such as peptides. The measured intensity strongly depends on preanalytical  and
environmental factors like sample collection, sample storing, room temperature,
air humidity, crystallization etc. \citep{Baggerly2004, Leek2010, Leichtle2013}.

Further confounders are introduced by so-called batch effects. These are systematical differences
that hide the true biological effect and that are caused by different experimental
conditions, for instance a different preanalytical processing, measurements on different
days by different operators in different laboratories on different devices
\citep{Hu2005, Leek2010, Gregori2012}.  

The systematic errors can be stronger than the real biological effect, and are
best minimized already at the stage of data acquisition by strictly adhering 
to a standardized preanalytical and experimental protocol \citep{Baggerly2004}.

Note that unlike in other omics data, such as gene expression data, batch effects and other systematic errors can be the source of 
shifts both on the $x$-axis (\ac{mz}
values) and on the $y$-axis (intensity values).  Hence, to ensure the validity of 
any subsequent statistical analysis, great care must be taken
to address both of these shift in preprocessing, by \emph{intensity calibration} (often called
normalization) and by \emph{peak alignment/warping} (see also \ref{subsec:pa}).

Methods to calibrate peak intensities can be divided into \emph{local} and
\emph{global} approaches \citep{Meuleman2008}.
In a \emph{local} calibration each single spectrum is calibrated on its own,
by matching a specified characteristic such as the median, the mean or the
\ac{TIC} \citep{Callister2006,Meuleman2008, Borgaonkar2010}.  In contrast,
\emph{global} approaches use information across multiple spectra, e.g. employing 
linear regression normalization \citep{Callister2006}, quantile normalization
\citep{Bolstad2003}, or \ac{PQN} \citep{Dieterle2006}.

\Mq{} supports two \emph{local} and one \emph{global} method. Specifically, it
implements the
\ac{TIC} and median calibration as well as \ac{PQN}.
In \ac{PQN} all spectra are calibrated using the \ac{TIC} calibration
first. Subsequently, a median reference spectrum is created and the intensities in 
all spectra are standardized using the reference spectrum and a 
spectrum-specific median is calculated for each spectrum. 
Finally,  each spectrum is rescaled by the median of the ratios of its intensity
values and that of the reference spectrum  \citep{Dieterle2006}.

It has been shown that applying intensity calibration is an essential step in preprocessing
\citep{Meuleman2008}. Despite
its simplicity \ac{TIC} is often the best choice, especially to account for
effects between technical replicates \citep{Shin2006, Meuleman2008}.

\subsection{Peak Detection}

Peak detection is a further step in processing mass spectrometry data,
serving both to identify potential relevant features as well as to reduce
the dimensionality of the data.

\begin{figure}[t!]
  \includegraphics[width=\textwidth]{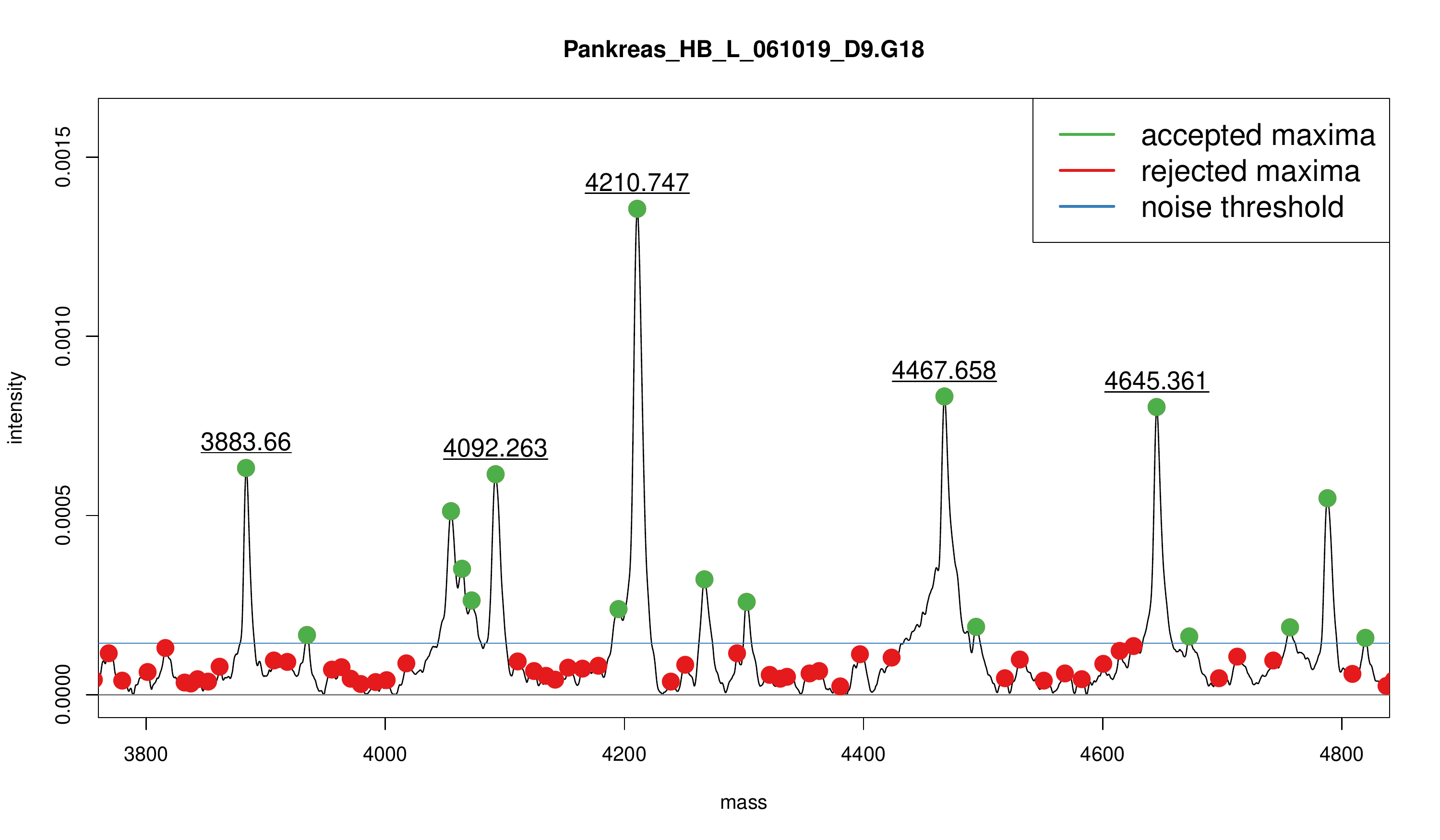}
  \caption[Peak Detection]{Detail view of a \ac{MALDITOF} spectrum from
    \citet{Fiedler2009}. Local maxima are marked with points (red: rejected
    maxima, green: peaks). The blue line represents the estimated noise baseline
    as estimated by \ac{MAD}.}
  \label{fig:pd}
\end{figure}

\Mq{} provides the most commonly used peak detection method based on finding local
maxima \citep{Yasui2003, Tibshirani2004, PROcess, Morris2005, Smith2006,
Tracy2008}.
First a window is moved across the spectra and local maxima are detected.
Subsequently these local maxima are compared against a noise baseline which is
estimated by the \ac{MAD} or alternatively Friedman's SuperSmoother \citep{Friedman1984}.
If a local maximum is above a given \ac{SNR} it is considered a peak, whereas local
maxima
below the \ac{SNR} threshold are discarded (Fig~\ref{fig:pd}).

Some authors advocate peak detection methods based on 
on \emph{wavelets} \citep{Du2006, Lange2006}.  These methods are implemented in
the Bioconductor R packages
\Rpackage{MassSpecWavelet} and \Rpackage{xcms} \citep{Smith2006} and thus
are readily available if needed.

\subsection{Peak Alignment}
\label{subsec:pa}

As already noted above in Section~\ref{subsec:ic} 
not only the intensities but also the \ac{mz} values differ across
spectra, as result of  the many possible sources
of variation in the acquisition of mass spectrometry data.
 Methods to recalibrate the \ac{mz} values of the spectra are
referred to as \emph{peak alignment} or \emph{warping}.

A simple approach is the \ac{COW} algorithm \citep{Veselkov2009, Morris2010, Wang2010}.
\ac{COW} is based on pairwise comparisons of spectra and maximizes the correlation
to find an optimal shift.  The advantage of this approach is that correlation is
fast to compute and with the use of a reference spectrum the method is also 
applicable to simultaneous alignment of multiple spectra.  However, in actual data
the location shifts are typically of a nonlinear nature \citep{He2011}, thus methods
based on global linear shifts will often be ineffective in achieving an optimal alignment.
A possible workaround is to divide the spectrum into several parts and perform local linear
alignment instead.

An alternative, and much more flexible approach, is  \ac{DTW} which is
based on dynamic programming \citep{Torgrip2003, Toppo2008, Clifford2009, Kim2011}.
\ac{DTW} is a pairwise alignment approach that is guaranteed to find the optimal alignment by
comparing  each point in the first spectrum to
every other point in the second spectrum, and optimizing a distance score.
Dynamic programming techniques are used to substantially shortcut computational
time by means of an underlying decision tree \citep[e.g.][]{Sakoe1978}.
Still, \ac{DTW} is a computationally very expensive algorithm that also requires
substantial computer memory, especially for multiple alignment.

As a compromise, recently the  \ac{PTW} approach has been suggested \citep{Jeffries2005, Lin2005,
Bloemberg2010, He2011, Wehrens2015} where a polynomial functions is used to
stretch or shrink a spectrum to increase the similarity between them. \ac{PTW} is a
very fast method that is also able to correct for
non-linear shifts. As in all previously mentioned method a
reference spectrum is necessary for multiple alignment. Note that the use of a reference
spectrum requires prior calibration of the intensity values \citep{Smith2013}.

Finally,  another simple strategy to align peaks is based on clustering respectively
creation of bins of similar \ac{mz} values \citep{Yasui2003a, Tibshirani2004,
Tracy2008}.  This is a fast and easy to implement approach, and  in contrast to \ac{DTW},
\ac{COW} and \ac{PTW} it offers the possibility to align all spectra simultaneously.
However, the clustering approach is valid only when there are relatively 
small shifts around the true peak
position, hence this approach is only applicable if there are only mild distortions
in the \ac{mz} values.

In \Mq{} we use nonlinear warping of peaks \citep{He2011,Wehrens2015}. First we align
the \ac{mz} values of the peaks using \ac{PTW} and subsequently we employ binning
to identify common peak positions across spectra. Note that in contrast to the standard version
of \ac{PTW} we work on peak level rather than on the whole spectrum.

\begin{figure}[t]
  \includegraphics[width=\textwidth]{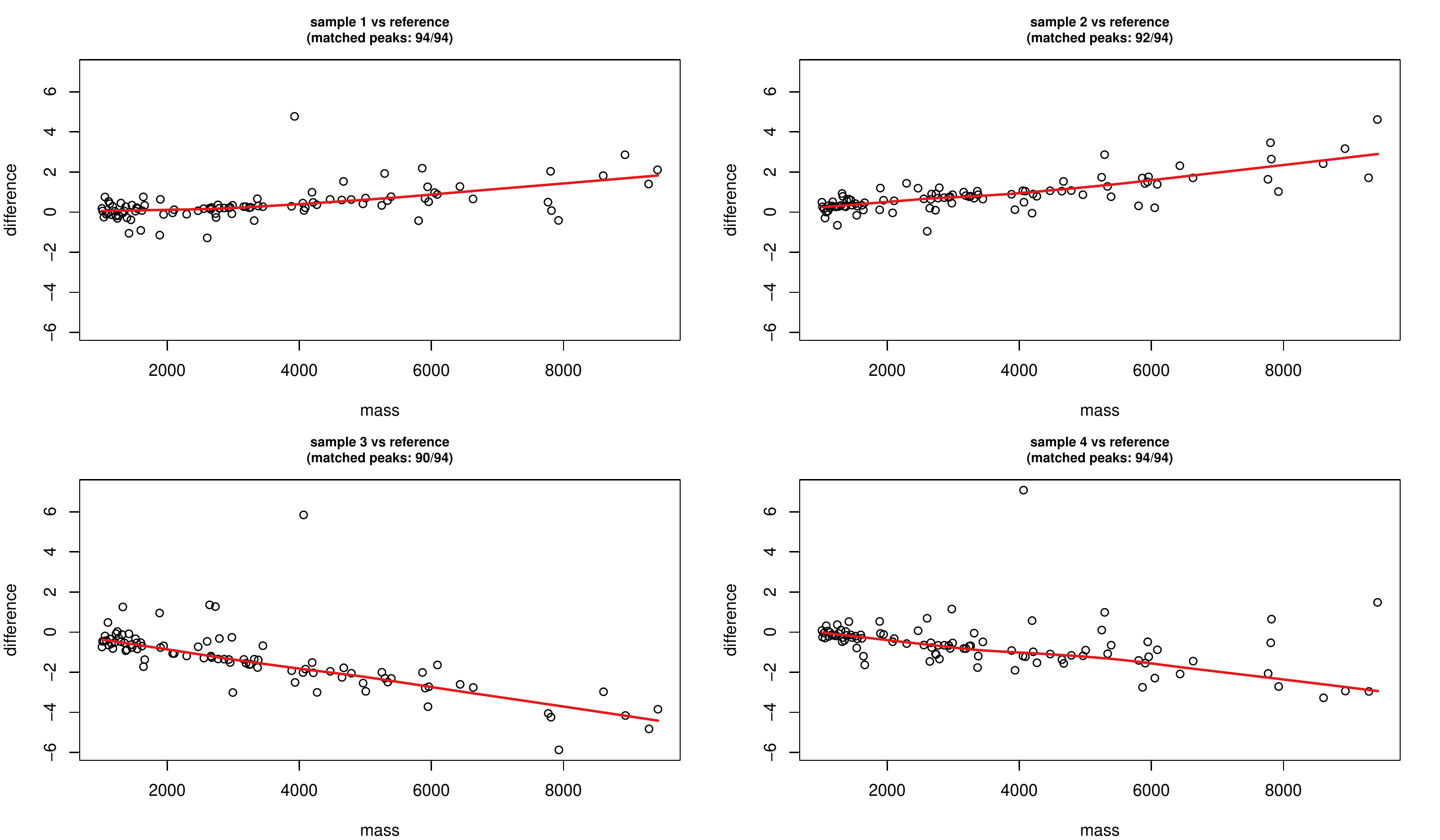}
  \caption[Warping Functions]{Example warping function for four different peak lists.
    The x-axis represents the peak position and the y-axis the
    difference from the reference peak list. The red line shows the calculated
    warping function. The number of matched peaks out of all reference peaks
    is also shown for each spectrum.
  }
  \label{fig:wp_fun}
\end{figure}

Our peak alignment algorithm in \Mq{} starts by looking for stable peaks, which
are defined as  high peaks in
defined, coarse \ac{mz} ranges that are present in most spectra. The \ac{mz} of
the peaks is averaged and used as reference peak list (also known as
anchor or landmark peaks --- see \citet{Wang2010}.
Next, \Mq{} computes a \ac{LOWESS} curve or polynomial-based function to warp the peaks
of each spectra against the reference peaks (Fig.~\ref{fig:wp_fun}).
As not all reference peaks are found in each spectrum, the number of
matched peaks out of all reference peaks is reported by \Mq\ for information. 

\begin{figure}[t]
  \includegraphics[width=\textwidth]{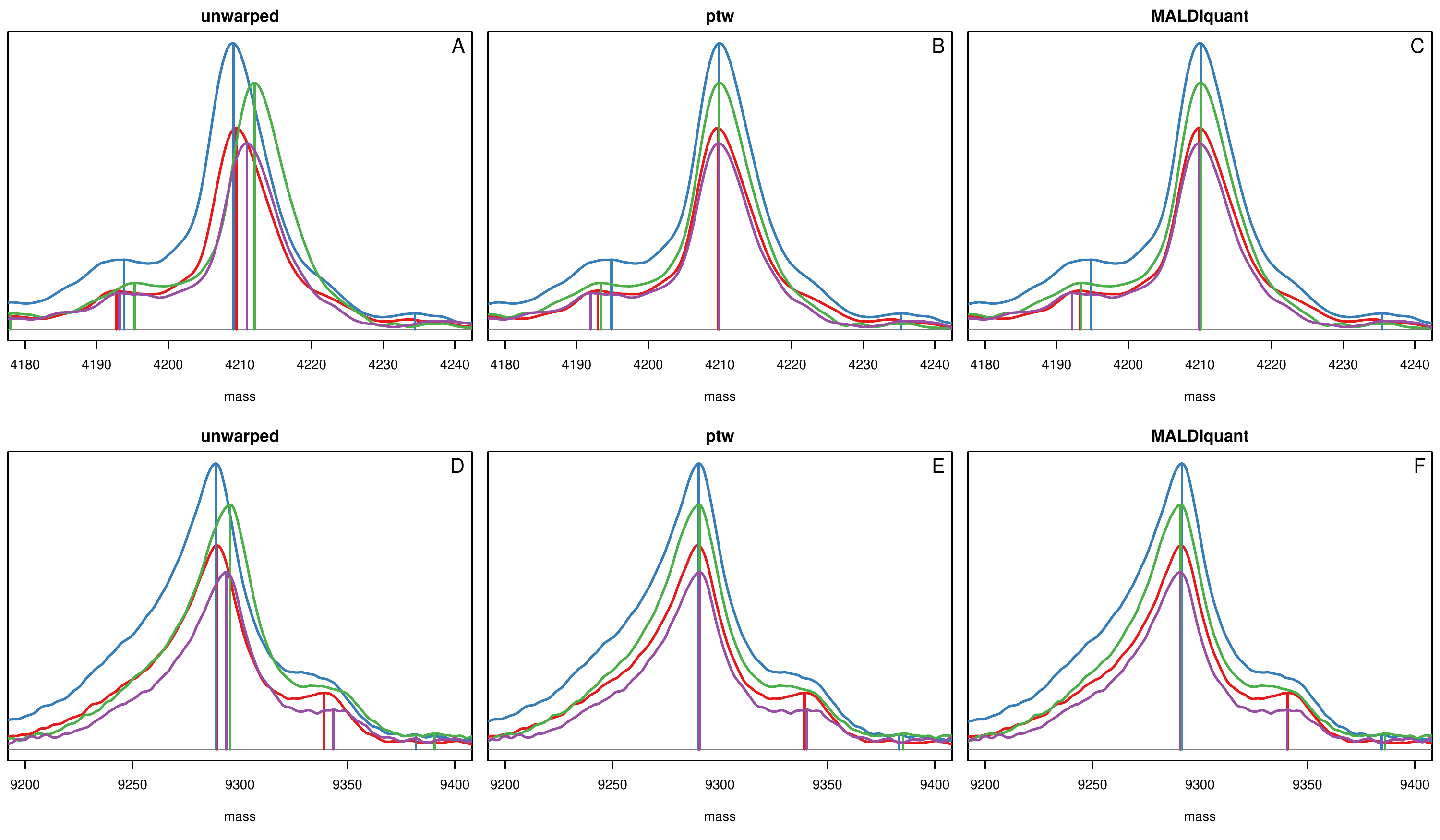}
  \caption[Comparison warping]{Comparison of two peaks (top row and bottom row)
    present in four
    \ac{MALDITOF} spectra from \citet{Fiedler2009}.
    (A, D) unaligned;
    (B, E) warped using the \ac{PTW} algorithm;
    (C, F) warped using \Mq{}'s peak based \ac{PTW}.}
  \label{fig:wp}
\end{figure}

Due to using the peaks instead of the whole spectral data the alignment 
approach implemented in \Mq{}
is much faster than traditional \ac{PTW}, still the results are
comparable (Fig.~\ref{fig:wp}).
Another important advantage of our approach is that only \ac{mz} values
are used for calibration,  which implies that our
approach does not require perfectly calibrated intensity values as is the case
for full spectrum-based alignment methods.

After performing alignment, peak positions of identical features across
spectra will become very similar but in general not
numerically identical. Thus, as final step grouping of \ac{mz} values into bins is
needed. For this purpose \Mq{} uses the following simple clustering algorithm:
The \ac{mz} values are sorted in ascending order and split recursively
at the largest gap until all \ac{mz} values in the resulting bins are from
different samples and their individual \ac{mz} values are in a small
user-defined tolerance range around their mean. The latter becomes the new
\ac{mz} value for all corresponding peaks in the associated bin.

\subsection{Subsequent Statistical Analysis}

With peak alignment the task of \Mq\ to transform raw mass spectrometry data
into a matrix containing intensity measurements of potentially useful  \ac{mz} values
is complete.

Subsequently, the resulting feature intensity matrix can be used with any preferred univariate or multivariate
analysis technique, e.g. to identify peaks that are useful for predicting a desired
outcome, or simply to rank features with regard to group separation \citep[e.g.][]{GibbStrimmer2015}.

In the following section we will describe in detail how such an analysis may be conducted.
For more examples please see the \Mq\ homepage.

\section{Case Study}

\subsection{Dataset}

For illustration how to use \Mq\ in practical data analysis 
we now show in detail how to use the software by application to the 
mass spectromety data published in \citet{Fiedler2009}. The aim of this study was to determine
proteomic biomarkers to discriminate patients with pancreas cancer from healthy persons.
As part of their study the authors collected serum samples of 40 patients with diagnosed pancreas
cancer as well as 40 healthy controls as training dataset. For each sample 4
technical replicates were obtained. These 320 samples were processed following a
standardized protocol for serum peptidomics and subsequently analyzed in a
linear \ac{MALDITOF} mass spectrometer. For details on the experimental setup
 we refer to the original study.

Half of the patients and controls were recruited at the University Hospital
Heidelberg and the University Hospital Leipzig.  Due to the presence of
strong batch effects we restrict ourselves to the samples from Heidelberg,
leading to a raw data set containing 160 spectra for 40 probands, of which 20 were
diagnosed with pancreatic cancer and 20 are healthy controls.
\citet{Fiedler2009} found marker peaks at \ac{mz} 3884 (double charged) and 7767
(single charged) and correspondingly suggested \ac{PF4} as
potential marker, arguing that \ac{PF4} is down-regulated in blood serum of
patients with pancreatic cancer.

\subsection{Preparations}

Prior to preprocessing the data we first need to set up our \R{} environment
by  install the
necessary packages, namely \Mq{} \citep{MALDIquant}, \MqF{}, \Rpackage{sda} and
\Rpackage{crossval}, and also download the data set:

\begin{knitrout}
\definecolor{shadecolor}{rgb}{0.969, 0.969, 0.969}\color{fgcolor}\begin{kframe}
\begin{alltt}
\hlkwd{install.packages}\hlstd{(}\hlkwd{c}\hlstd{(}\hlstr{"MALDIquant"}\hlstd{,} \hlstr{"MALDIquantForeign"}\hlstd{,}
                   \hlstr{"sda"}\hlstd{,} \hlstr{"crossval"}\hlstd{))}
\end{alltt}
\end{kframe}
\end{knitrout}

\begin{knitrout}
\definecolor{shadecolor}{rgb}{0.969, 0.969, 0.969}\color{fgcolor}\begin{kframe}
\begin{alltt}
\hlcom{## load packages}
\hlkwd{library}\hlstd{(}\hlstr{"MALDIquant"}\hlstd{)}
\end{alltt}

{\ttfamily\noindent\itshape\color{messagecolor}{Loading required package: methods

This is MALDIquant version 1.13\\Quantitative Analysis of Mass Spectrometry Data\\ See '?MALDIquant' for more information about this package.}}\begin{alltt}
\hlkwd{library}\hlstd{(}\hlstr{"MALDIquantForeign"}\hlstd{)}

\hlcom{## download the raw spectra data (approx. 90 MB)}
\hlstd{githubUrl} \hlkwb{<-} \hlkwd{paste0}\hlstd{(}\hlstr{"https://raw.githubusercontent.com/sgibb/"}\hlstd{,}
                    \hlstr{"MALDIquantExamples/master/inst/extdata/"}\hlstd{,}
                    \hlstr{"fiedler2009/"}\hlstd{)}
\hlstd{downloader}\hlopt{::}\hlkwd{download}\hlstd{(}\hlkwd{paste0}\hlstd{(githubUrl,} \hlstr{"spectra.tar.gz"}\hlstd{),}
                     \hlstr{"fiedler2009spectra.tar.gz"}\hlstd{)}
\hlcom{## download metadata}
\hlstd{downloader}\hlopt{::}\hlkwd{download}\hlstd{(}\hlkwd{paste0}\hlstd{(githubUrl,} \hlstr{"spectra_info.csv"}\hlstd{),}
                     \hlstr{"fiedler2009info.csv"}\hlstd{)}
\end{alltt}
\end{kframe}
\end{knitrout}

\subsection{Import Raw Data and Quality Control}

The first step in the analysis comprises importing the raw data into the \R{} environment.
As the raw data set contains both the samples from Heidelberg and Leipzig we filter out the
samples from Leipzig, so that our final data set only contains the Heidelberg patients and controls:
\begin{knitrout}
\definecolor{shadecolor}{rgb}{0.969, 0.969, 0.969}\color{fgcolor}\begin{kframe}
\begin{alltt}
\hlcom{## import the spectra}
\hlstd{spectra} \hlkwb{<-} \hlkwd{import}\hlstd{(}\hlstr{"fiedler2009spectra.tar.gz"}\hlstd{,} \hlkwc{verbose}\hlstd{=}\hlnum{FALSE}\hlstd{)}

\hlcom{## import metadata}
\hlstd{spectra.info} \hlkwb{<-} \hlkwd{read.csv}\hlstd{(}\hlstr{"fiedler2009info.csv"}\hlstd{)}

\hlcom{## keep data from Heidelberg}
\hlstd{isHeidelberg} \hlkwb{<-} \hlstd{spectra.info}\hlopt{$}\hlstd{location} \hlopt{==} \hlstr{"heidelberg"}

\hlstd{spectra} \hlkwb{<-} \hlstd{spectra[isHeidelberg]}
\hlstd{spectra.info} \hlkwb{<-} \hlstd{spectra.info[isHeidelberg,]}
\end{alltt}
\end{kframe}
\end{knitrout}

After importing the raw data it is recommend to perform basic sanity checks for quality control.
Below, we test whether all spectra contain the
same number of data points, are not empty and are regular, i.e. whether the 
differences between subsequent \ac{mz}
values are constant:
\begin{knitrout}
\definecolor{shadecolor}{rgb}{0.969, 0.969, 0.969}\color{fgcolor}\begin{kframe}
\begin{alltt}
\hlkwd{table}\hlstd{(}\hlkwd{lengths}\hlstd{(spectra))}
\end{alltt}
\begin{verbatim}

42388 
  160 
\end{verbatim}
\begin{alltt}
\hlkwd{any}\hlstd{(}\hlkwd{sapply}\hlstd{(spectra, isEmpty))}
\end{alltt}
\begin{verbatim}
[1] FALSE
\end{verbatim}
\begin{alltt}
\hlkwd{all}\hlstd{(}\hlkwd{sapply}\hlstd{(spectra, isRegular))}
\end{alltt}
\begin{verbatim}
[1] TRUE
\end{verbatim}
\end{kframe}
\end{knitrout}
Next, we ensure that all spectra cover the same \ac{mz} range. The `trim`
function automatically determines a suitable common \ac{mz} range if it is called without
any additional arguments:

\begin{knitrout}
\definecolor{shadecolor}{rgb}{0.969, 0.969, 0.969}\color{fgcolor}\begin{kframe}
\begin{alltt}
\hlstd{spectra} \hlkwb{<-} \hlkwd{trim}\hlstd{(spectra)}
\end{alltt}
\end{kframe}
\end{knitrout}
Finally, it is advised to inspect the spectra visually to discover any obviously distorted
measurements. Here, for reasons of space we only plot a single spectrum:

\begin{knitrout}
\definecolor{shadecolor}{rgb}{0.969, 0.969, 0.969}\color{fgcolor}\begin{kframe}
\begin{alltt}
\hlkwd{plot}\hlstd{(spectra[[}\hlnum{47}\hlstd{]],} \hlkwc{sub}\hlstd{=}\hlstr{""}\hlstd{)}
\end{alltt}
\end{kframe}\begin{figure}

{\centering \includegraphics[width=\maxwidth]{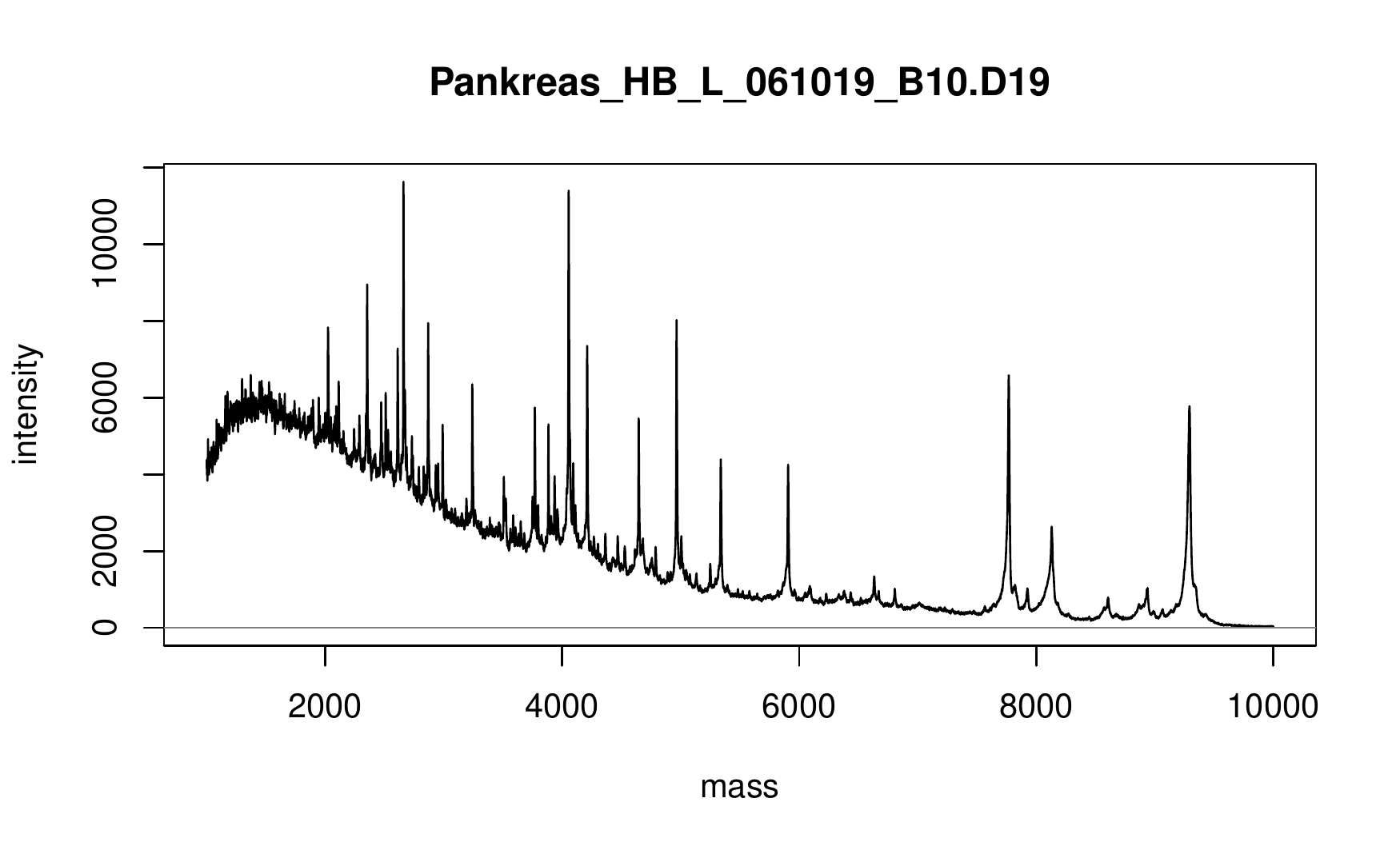} 

}

\caption[Example of a raw, uncalibrated mass spectrum]{Example of a raw, uncalibrated mass spectrum.}\label{fig:plot}
\end{figure}

\end{knitrout}

\subsection{Transformation and Smoothing}

Next, we perform variance stabilization by applying the square root transformation
to the raw data, and subsequently use a  41 point \emph{Savitzky-Golay}-Filter
\citep{Savitzky1964} to smooth the spectra:
\begin{knitrout}
\definecolor{shadecolor}{rgb}{0.969, 0.969, 0.969}\color{fgcolor}\begin{kframe}
\begin{alltt}
\hlstd{spectra} \hlkwb{<-} \hlkwd{transformIntensity}\hlstd{(spectra,} \hlkwc{method}\hlstd{=}\hlstr{"sqrt"}\hlstd{)}

\hlstd{spectra} \hlkwb{<-} \hlkwd{smoothIntensity}\hlstd{(spectra,} \hlkwc{method}\hlstd{=}\hlstr{"SavitzkyGolay"}\hlstd{,}
                           \hlkwc{halfWindowSize}\hlstd{=}\hlnum{20}\hlstd{)}
\end{alltt}
\end{kframe}
\end{knitrout}

\subsection{Baseline Correction}

In the next step we address the problem of matrix-effects and chemical noise that 
result in an elevated baseline. In our analysis we use 
the \emph{SNIP} algorithm \citep{Ryan1988} to estimate the baseline for each spectrum. 
Subsequently, the estimated baseline is subtracted to yield baseline-adjusted 
spectra:

\begin{knitrout}
\definecolor{shadecolor}{rgb}{0.969, 0.969, 0.969}\color{fgcolor}\begin{kframe}
\begin{alltt}
\hlstd{baseline} \hlkwb{<-} \hlkwd{estimateBaseline}\hlstd{(spectra[[}\hlnum{1}\hlstd{]],} \hlkwc{method}\hlstd{=}\hlstr{"SNIP"}\hlstd{,}
                             \hlkwc{iterations}\hlstd{=}\hlnum{150}\hlstd{)}
\hlkwd{plot}\hlstd{(spectra[[}\hlnum{1}\hlstd{]],} \hlkwc{sub}\hlstd{=}\hlstr{""}\hlstd{)}
\hlkwd{lines}\hlstd{(baseline,} \hlkwc{col}\hlstd{=}\hlstr{"red"}\hlstd{,} \hlkwc{lwd}\hlstd{=}\hlnum{2}\hlstd{)}
\end{alltt}
\end{kframe}\begin{figure}

{\centering \includegraphics[width=\maxwidth]{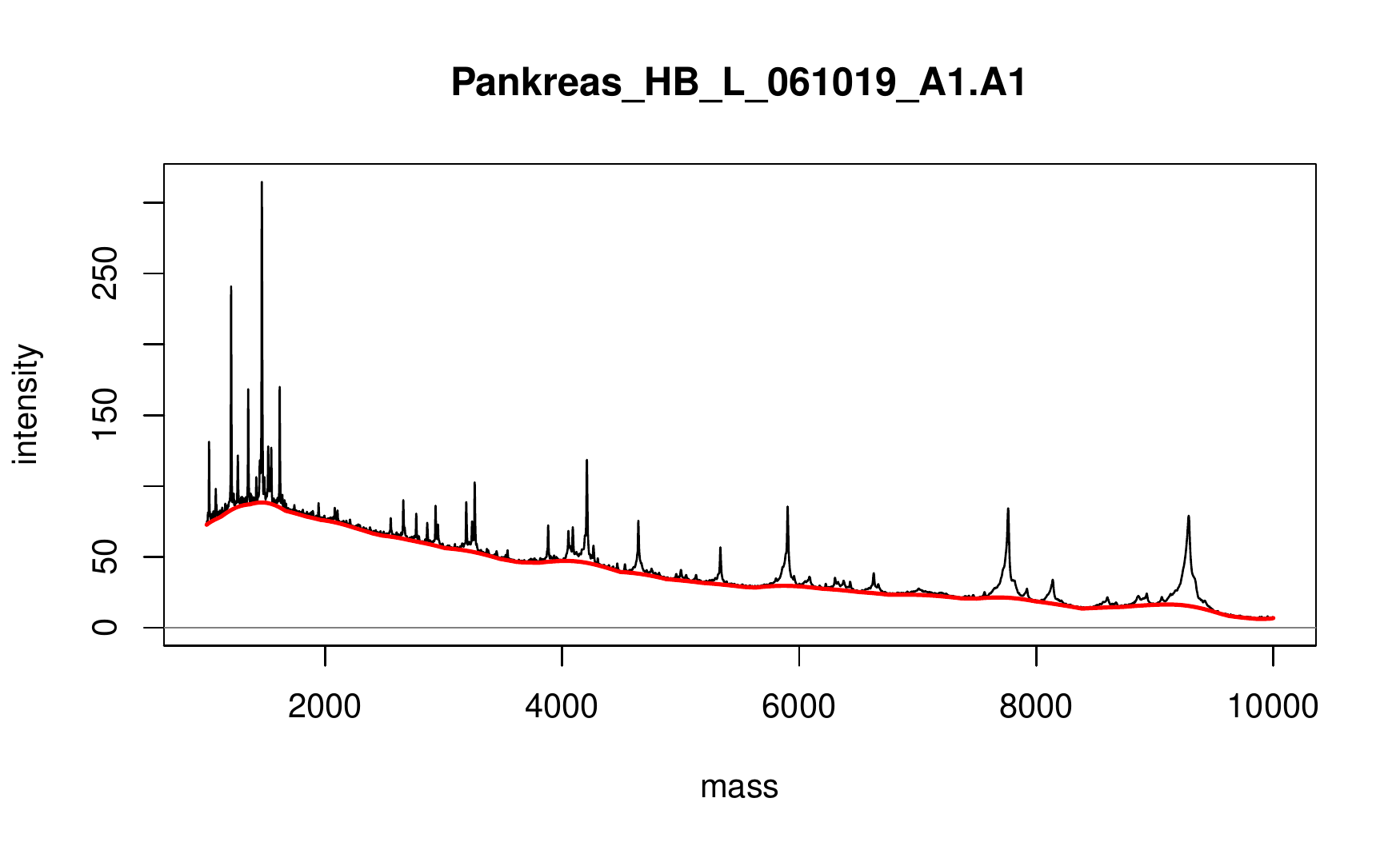} 

}

\caption[Baseline estimated using the SNIP method (red line)]{Baseline estimated using the SNIP method (red line).}\label{fig:bes}
\end{figure}

\end{knitrout}

\begin{knitrout}
\definecolor{shadecolor}{rgb}{0.969, 0.969, 0.969}\color{fgcolor}\begin{kframe}
\begin{alltt}
\hlstd{spectra} \hlkwb{<-} \hlkwd{removeBaseline}\hlstd{(spectra,} \hlkwc{method}\hlstd{=}\hlstr{"SNIP"}\hlstd{,}
                          \hlkwc{iterations}\hlstd{=}\hlnum{150}\hlstd{)}
\hlkwd{plot}\hlstd{(spectra[[}\hlnum{1}\hlstd{]],} \hlkwc{sub}\hlstd{=}\hlstr{""}\hlstd{)}
\end{alltt}
\end{kframe}\begin{figure}

{\centering \includegraphics[width=\maxwidth]{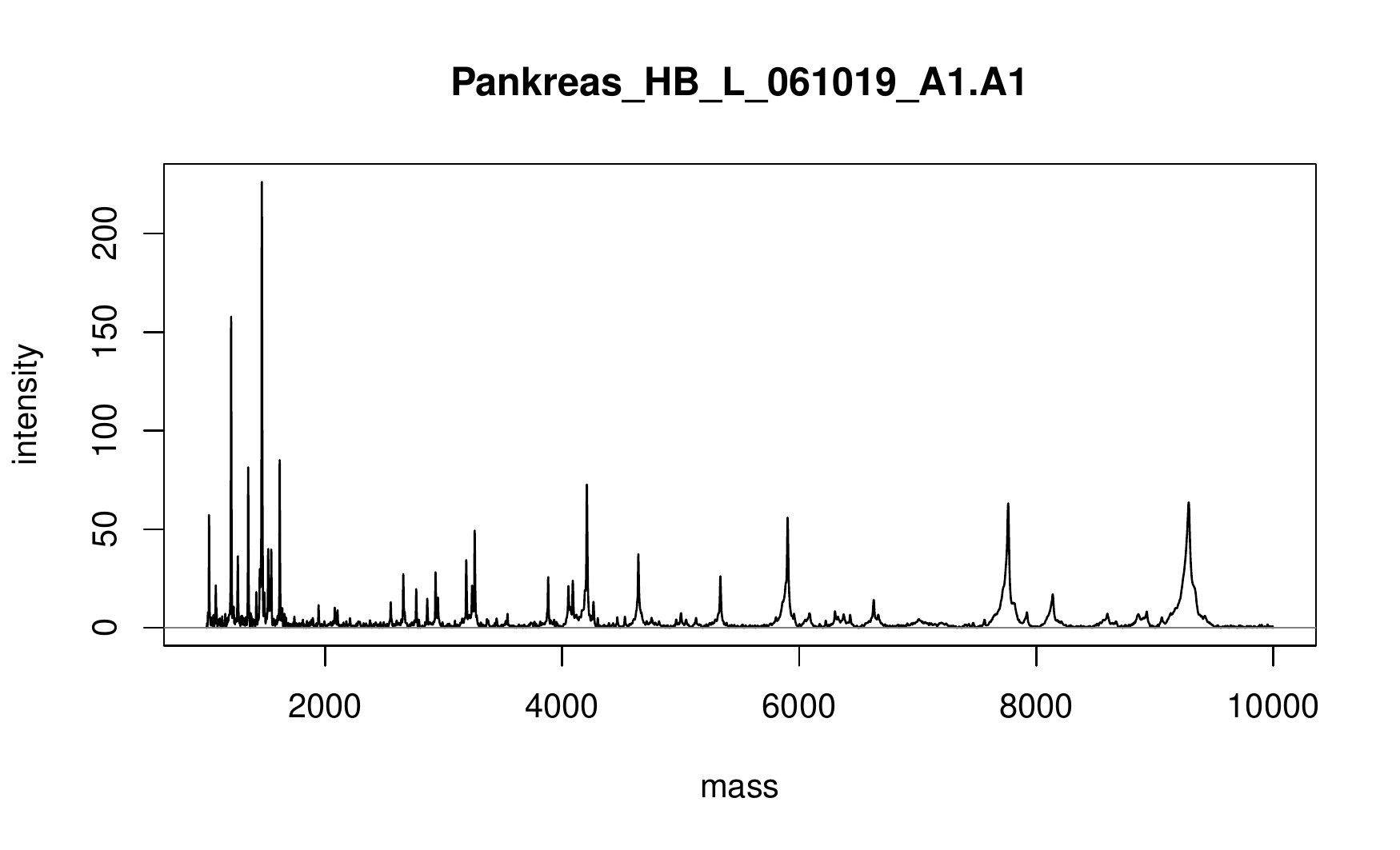} 

}

\caption[Mass spectrum after baseline correction]{Mass spectrum after baseline correction.}\label{fig:bc}
\end{figure}

\end{knitrout}

\subsection{Intensity Calibration and Alignment}

After baseline correction we  calibrate each spectrum by equalizing
the \ac{TIC} across spectra.
After normalizing the intensities we also  need to adjust the mass values. 
This is done by the peak
based warping algorithm implemented in \Mq{}.  In the example code
the function \Rfunction{alignSpectra} 
acts as a simple wrapper
around more complicated procedures. For a finer control of the underlying
procedures the function
\Rfunction{determineWarpingFunctions} may be used alternatively:

\begin{knitrout}
\definecolor{shadecolor}{rgb}{0.969, 0.969, 0.969}\color{fgcolor}\begin{kframe}
\begin{alltt}
\hlstd{spectra} \hlkwb{<-} \hlkwd{calibrateIntensity}\hlstd{(spectra,} \hlkwc{method}\hlstd{=}\hlstr{"TIC"}\hlstd{)}
\hlstd{spectra} \hlkwb{<-} \hlkwd{alignSpectra}\hlstd{(spectra)}
\end{alltt}
\end{kframe}
\end{knitrout}

Next, we average the technical replicates before we search for peaks and update
our meta information accordingly:

\begin{knitrout}
\definecolor{shadecolor}{rgb}{0.969, 0.969, 0.969}\color{fgcolor}\begin{kframe}
\begin{alltt}
\hlstd{avgSpectra} \hlkwb{<-}
  \hlkwd{averageMassSpectra}\hlstd{(spectra,} \hlkwc{labels}\hlstd{=spectra.info}\hlopt{$}\hlstd{patientID)}
\hlstd{avgSpectra.info} \hlkwb{<-}
  \hlstd{spectra.info[}\hlopt{!}\hlkwd{duplicated}\hlstd{(spectra.info}\hlopt{$}\hlstd{patientID), ]}
\end{alltt}
\end{kframe}
\end{knitrout}

\subsection{Peak Detection and Computation of Intensity Matrix}

Peak detection is the crucial step to identify features and to
reduce the dimensionality of the data. Before performing 
peak detection we first estimate the noise of selected spectra 
to investigate suitable settings for the \emph{signal-to-noise ratio} (SNR):
\begin{knitrout}
\definecolor{shadecolor}{rgb}{0.969, 0.969, 0.969}\color{fgcolor}\begin{kframe}
\begin{alltt}
\hlstd{noise} \hlkwb{<-} \hlkwd{estimateNoise}\hlstd{(avgSpectra[[}\hlnum{1}\hlstd{]])}
\hlkwd{plot}\hlstd{(avgSpectra[[}\hlnum{1}\hlstd{]],} \hlkwc{xlim}\hlstd{=}\hlkwd{c}\hlstd{(}\hlnum{4000}\hlstd{,} \hlnum{5000}\hlstd{),} \hlkwc{ylim}\hlstd{=}\hlkwd{c}\hlstd{(}\hlnum{0}\hlstd{,} \hlnum{0.002}\hlstd{))}
\hlkwd{lines}\hlstd{(noise,} \hlkwc{col}\hlstd{=}\hlstr{"red"}\hlstd{)}                     \hlcom{# SNR == 1}
\hlkwd{lines}\hlstd{(noise[,} \hlnum{1}\hlstd{],} \hlnum{2}\hlopt{*}\hlstd{noise[,} \hlnum{2}\hlstd{],} \hlkwc{col}\hlstd{=}\hlstr{"blue"}\hlstd{)} \hlcom{# SNR == 2}
\end{alltt}
\end{kframe}\begin{figure}

{\centering \includegraphics[width=\maxwidth]{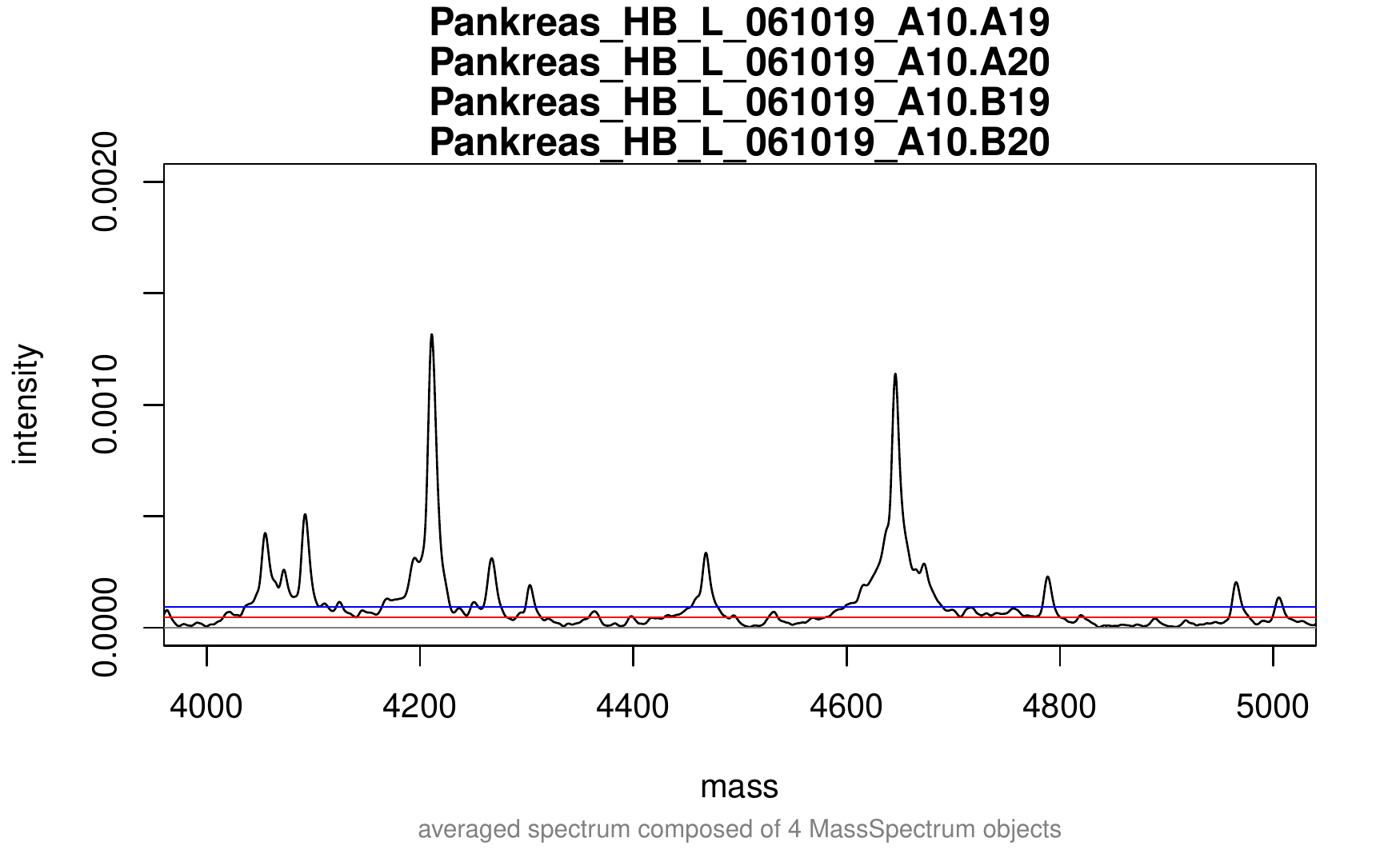} 

}

\caption[Thresholds based on signal to noise ratio (SNR)]{Thresholds based on signal to noise ratio (SNR): SNR=1 (red line) and SNR=2 (blue line).}\label{fig:noise}
\end{figure}

\end{knitrout}

In this case we decide to set a \emph{SNR} of 2 (blue line)
and then run the peak detection algorithm:

\begin{knitrout}
\definecolor{shadecolor}{rgb}{0.969, 0.969, 0.969}\color{fgcolor}\begin{kframe}
\begin{alltt}
\hlstd{peaks} \hlkwb{<-} \hlkwd{detectPeaks}\hlstd{(avgSpectra,} \hlkwc{SNR}\hlstd{=}\hlnum{2}\hlstd{,} \hlkwc{halfWindowSize}\hlstd{=}\hlnum{20}\hlstd{)}
\end{alltt}
\end{kframe}
\end{knitrout}

\begin{knitrout}
\definecolor{shadecolor}{rgb}{0.969, 0.969, 0.969}\color{fgcolor}\begin{kframe}
\begin{alltt}
\hlkwd{plot}\hlstd{(avgSpectra[[}\hlnum{1}\hlstd{]],} \hlkwc{xlim}\hlstd{=}\hlkwd{c}\hlstd{(}\hlnum{4000}\hlstd{,} \hlnum{5000}\hlstd{),} \hlkwc{ylim}\hlstd{=}\hlkwd{c}\hlstd{(}\hlnum{0}\hlstd{,} \hlnum{0.002}\hlstd{))}
\hlkwd{points}\hlstd{(peaks[[}\hlnum{1}\hlstd{]],} \hlkwc{col}\hlstd{=}\hlstr{"red"}\hlstd{,} \hlkwc{pch}\hlstd{=}\hlnum{4}\hlstd{)}
\end{alltt}
\end{kframe}\begin{figure}

{\centering \includegraphics[width=\maxwidth]{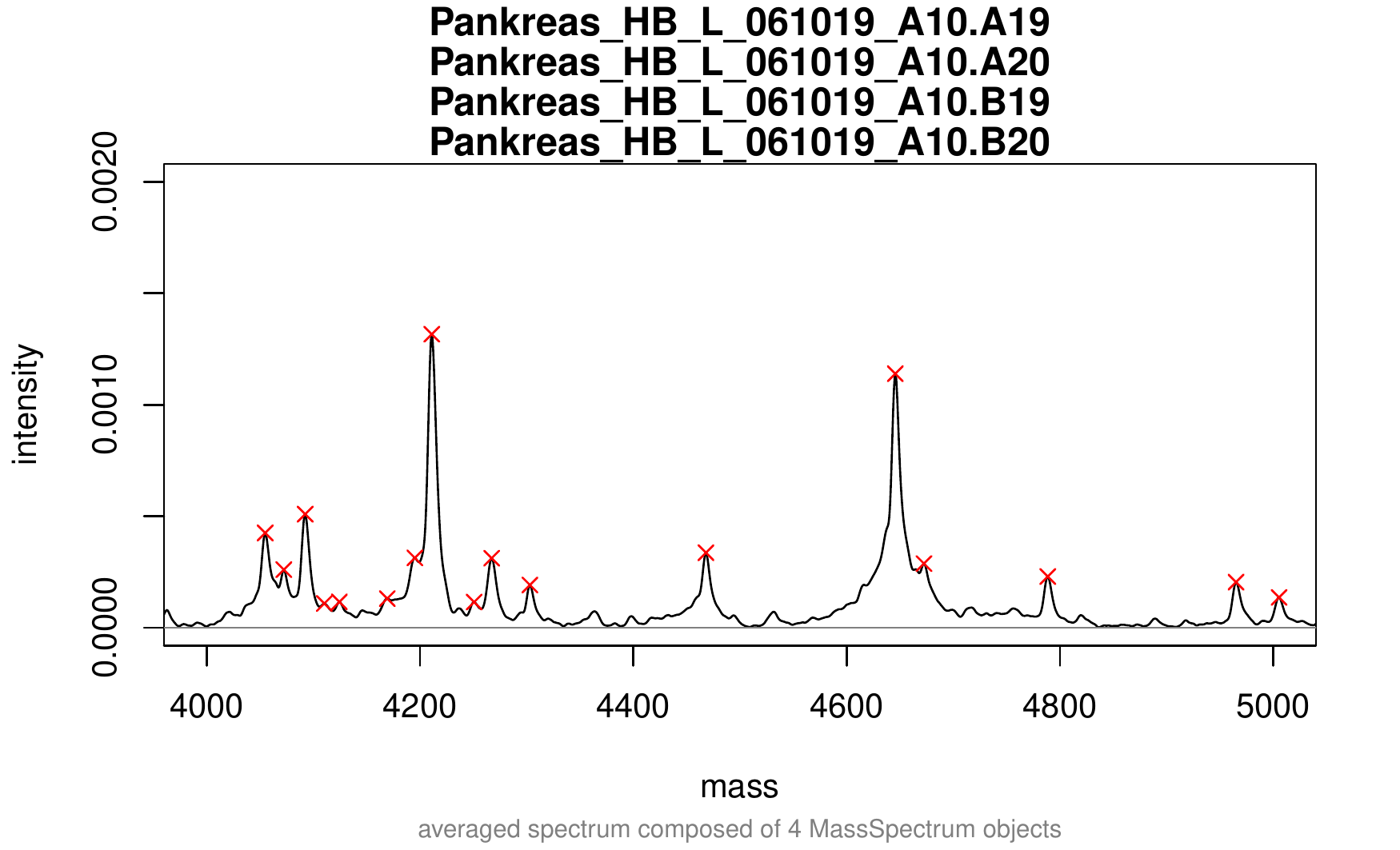} 

}

\caption[Peaks indentified above SNR=2 threshold]{Peaks indentified above SNR=2 threshold.}\label{fig:pdp}
\end{figure}

\end{knitrout}

After the alignment the peak positions (mass) are very similar but not numerically
identical. Consequently, binning is required to achieve identity:

\begin{knitrout}
\definecolor{shadecolor}{rgb}{0.969, 0.969, 0.969}\color{fgcolor}\begin{kframe}
\begin{alltt}
\hlstd{peaks} \hlkwb{<-} \hlkwd{binPeaks}\hlstd{(peaks)}
\end{alltt}
\end{kframe}
\end{knitrout}

In peak detection we choose a very low signal-to-noise ratio to keep as many features as possible.
Using the information about class labels we can now filter out false positive peaks,
by  removing peaks that appear in less than 50
\% of all spectra in each group:

\begin{knitrout}
\definecolor{shadecolor}{rgb}{0.969, 0.969, 0.969}\color{fgcolor}\begin{kframe}
\begin{alltt}
\hlstd{peaks} \hlkwb{<-} \hlkwd{filterPeaks}\hlstd{(peaks,} \hlkwc{minFrequency}\hlstd{=}\hlkwd{c}\hlstd{(}\hlnum{0.5}\hlstd{,} \hlnum{0.5}\hlstd{),}
                     \hlkwc{labels}\hlstd{=avgSpectra.info}\hlopt{$}\hlstd{health,}
                     \hlkwc{mergeWhitelists}\hlstd{=}\hlnum{TRUE}\hlstd{)}
\end{alltt}
\end{kframe}
\end{knitrout}

As final step in \Mq\ we create the feature intensity matrix, and for convenience
label the rows with the corresponding
patient ID:

\begin{knitrout}
\definecolor{shadecolor}{rgb}{0.969, 0.969, 0.969}\color{fgcolor}\begin{kframe}
\begin{alltt}
\hlstd{featureMatrix} \hlkwb{<-} \hlkwd{intensityMatrix}\hlstd{(peaks, avgSpectra)}
\hlkwd{rownames}\hlstd{(featureMatrix)} \hlkwb{<-} \hlstd{avgSpectra.info}\hlopt{$}\hlstd{patientID}
\hlkwd{dim}\hlstd{(featureMatrix)}
\end{alltt}
\begin{verbatim}
[1]  40 166
\end{verbatim}
\end{kframe}
\end{knitrout}

This matrix is the final output of \Mq\ and contains the calibrated intensity values for identified features across all 
spectra. It forms the basis for  higher-level statistical analysis.

\subsection{Feature Ranking and Classification}

The \citet{Fiedler2009} data set contains class labels for each spectrum (healthy versus cancer),
hence it is natural to perform a standard classification and feature ranking analysis.
A commonly used approach is Fisher's linear discriminant analysis (LDA), see 
\citet{Mertens2006} for details and applications to mass spectrometry analysis.
Many other classification approaches may also be applied, such as based on Random Forests
\citep{Breiman2001} or
peak discretization \citep{GibbStrimmer2015}.

Here, we use a variant of LDA implemented in
the R package \Rpackage{sda}  \citep{sda}.  In particular, we use diagonal discriminant analysis (DDA),
a special case of LDA with the assumption that the correlation among features (peaks) is negligible.
Despite this simplification this approach to classification is 
very effective, especially in high dimensions \citep{THNC03}.
In order to identify the most important class discriminating peaks
we use standard $t$-scores, which are the natural variable importance measure in DDA.

As a first step in our analysis we therefore compute the ranking of features by $t$-scores, and list the 10 top-ranking
features in Table~11.1:

\begin{knitrout}
\definecolor{shadecolor}{rgb}{0.969, 0.969, 0.969}\color{fgcolor}\begin{kframe}
\begin{alltt}
\hlkwd{library}\hlstd{(}\hlstr{"sda"}\hlstd{)}
\hlkwd{colnames}\hlstd{(featureMatrix)} \hlkwb{<-}
   \hlkwd{round}\hlstd{(}\hlkwd{as.double}\hlstd{(}\hlkwd{colnames}\hlstd{(featureMatrix)),}\hlnum{2}\hlstd{)}
\hlstd{Ytrain} \hlkwb{<-} \hlstd{avgSpectra.info}\hlopt{$}\hlstd{health}
\hlstd{ddar} \hlkwb{<-} \hlkwd{sda.ranking}\hlstd{(}\hlkwc{Xtrain}\hlstd{=featureMatrix,} \hlkwc{L}\hlstd{=Ytrain,} \hlkwc{fdr}\hlstd{=}\hlnum{FALSE}\hlstd{,}
                    \hlkwc{diagonal}\hlstd{=}\hlnum{TRUE}\hlstd{,} \hlkwc{verbose}\hlstd{=}\hlnum{FALSE}\hlstd{)}
\end{alltt}
\end{kframe}
\end{knitrout}
\begin{table}[ht]
\centering
\begin{tabular}{rrrrr}
  \hline
 & idx & score & t.cancer & t.control \\ 
  \hline
8936.97 & 158.00 & 90.69 & 9.52 & -9.52 \\ 
  4468.07 & 116.00 & 80.80 & 8.99 & -8.99 \\ 
  8868.27 & 157.00 & 80.06 & 8.95 & -8.95 \\ 
  4494.8 & 117.00 & 67.00 & 8.19 & -8.19 \\ 
  8989.2 & 159.00 & 66.19 & 8.14 & -8.14 \\ 
  5864.49 & 135.00 & 37.56 & -6.13 & 6.13 \\ 
  5906.17 & 136.00 & 34.43 & -5.87 & 5.87 \\ 
  2022.94 & 49.00 & 33.30 & 5.77 & -5.77 \\ 
  5945.57 & 137.00 & 32.66 & -5.71 & 5.71 \\ 
  1866.17 & 44.00 & 32.12 & 5.67 & -5.67 \\ 
   \hline
\end{tabular}
\caption{The ten top-ranking peaks as identified in the analysis.} 
\end{table}

To illustrate that feature selection based on the above feature ranking is 
indeed beneficial for subsequent analysis we 
apply
hierarchical cluster analysis based on the euclidean distance first to
the data set containing all features:

\begin{knitrout}
\definecolor{shadecolor}{rgb}{0.969, 0.969, 0.969}\color{fgcolor}\begin{kframe}
\begin{alltt}
\hlstd{distanceMatrix} \hlkwb{<-} \hlkwd{dist}\hlstd{(featureMatrix,} \hlkwc{method}\hlstd{=}\hlstr{"euclidean"}\hlstd{)}

\hlstd{hClust} \hlkwb{<-} \hlkwd{hclust}\hlstd{(distanceMatrix,} \hlkwc{method}\hlstd{=}\hlstr{"complete"}\hlstd{)}

\hlkwd{plot}\hlstd{(hClust,} \hlkwc{hang}\hlstd{=}\hlopt{-}\hlnum{1}\hlstd{)}
\end{alltt}
\end{kframe}\begin{figure}

{\centering \includegraphics[width=\maxwidth]{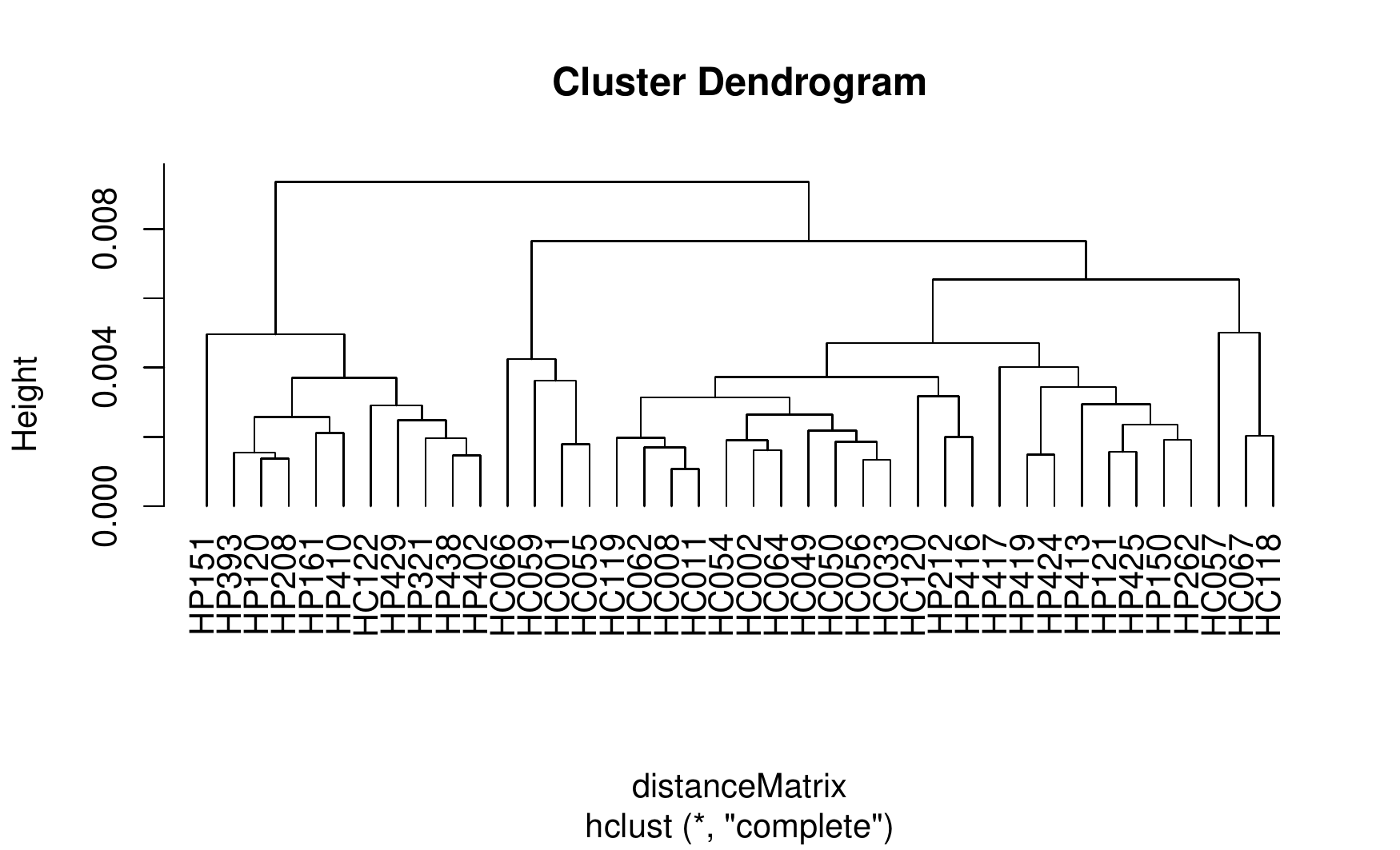} 

}

\caption[Hierarchical clustering of patient samples using all features]{Hierarchical clustering of patient samples using all features.}\label{fig:hclust}
\end{figure}

\end{knitrout}

Next, we repeat the above clustering on the data set containing 
only the best two top-ranking peaks:

\begin{knitrout}
\definecolor{shadecolor}{rgb}{0.969, 0.969, 0.969}\color{fgcolor}\begin{kframe}
\begin{alltt}
\hlstd{top} \hlkwb{<-} \hlstd{ddar[}\hlnum{1}\hlopt{:}\hlnum{2}\hlstd{,} \hlstr{"idx"}\hlstd{]}

\hlstd{distanceMatrixTop} \hlkwb{<-} \hlkwd{dist}\hlstd{(featureMatrix[, top],}
                          \hlkwc{method}\hlstd{=}\hlstr{"euclidean"}\hlstd{)}

\hlstd{hClustTop} \hlkwb{<-} \hlkwd{hclust}\hlstd{(distanceMatrixTop,} \hlkwc{method}\hlstd{=}\hlstr{"complete"}\hlstd{)}

\hlkwd{plot}\hlstd{(hClustTop,} \hlkwc{hang}\hlstd{=}\hlopt{-}\hlnum{1}\hlstd{)}
\end{alltt}
\end{kframe}\begin{figure}

{\centering \includegraphics[width=\maxwidth]{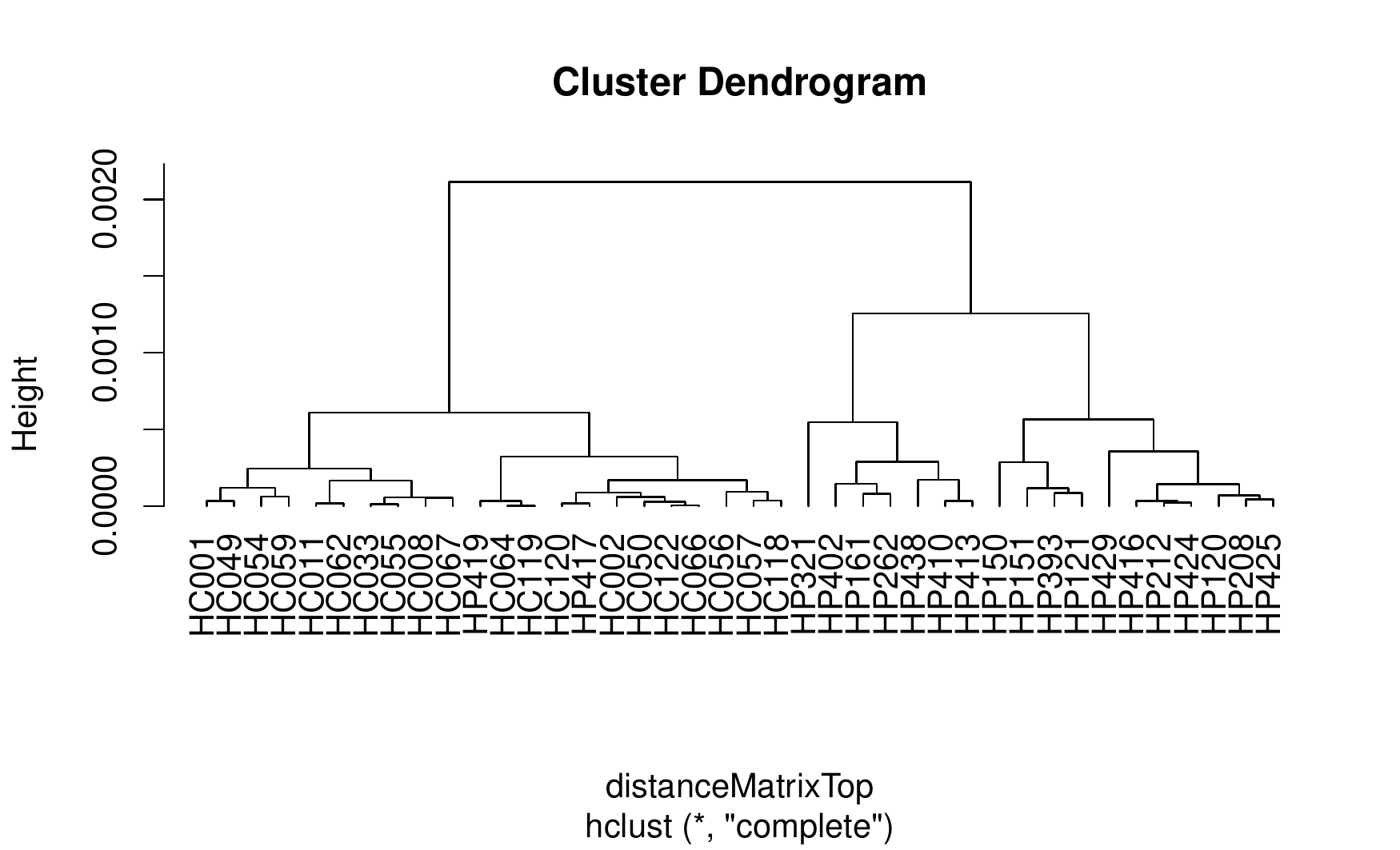} 

}

\caption[Hierarchical clustering of patient samples using only the best two top-ranking peaks]{Hierarchical clustering of patient samples using only the best two top-ranking peaks.}\label{fig:hclustfs}
\end{figure}

\end{knitrout}

As can be seen by comparison of the 
two trees, as a result of the feature selection we obtain a nearly
perfect split between the Heidelberg pancreas cancer samples (labeled "HP") and 
the Heidelberg control group (labeled "HC").

The strong predictive capabilities of the first two discovered peaks can be
further quantified by conducting a cross-validation analysis to estimate the prediction
error.  We use the \Rpackage{crossval} \citep{crossval} package to perform
a 10-fold cross validation using the predictor containing only  the two selected peaks:

\begin{knitrout}
\definecolor{shadecolor}{rgb}{0.969, 0.969, 0.969}\color{fgcolor}\begin{kframe}
\begin{alltt}
\hlkwd{library}\hlstd{(}\hlstr{"crossval"}\hlstd{)}
\hlcom{# create a prediction function for the cross validation}
\hlstd{predfun.dda} \hlkwb{<-} \hlkwa{function}\hlstd{(}\hlkwc{Xtrain}\hlstd{,} \hlkwc{Ytrain}\hlstd{,} \hlkwc{Xtest}\hlstd{,} \hlkwc{Ytest}\hlstd{,}
                        \hlkwc{negative}\hlstd{) \{}
  \hlstd{dda.fit} \hlkwb{<-} \hlkwd{sda}\hlstd{(Xtrain, Ytrain,} \hlkwc{diagonal}\hlstd{=}\hlnum{TRUE}\hlstd{,} \hlkwc{verbose}\hlstd{=}\hlnum{FALSE}\hlstd{)}
  \hlstd{ynew} \hlkwb{<-} \hlkwd{predict}\hlstd{(dda.fit, Xtest,} \hlkwc{verbose}\hlstd{=}\hlnum{FALSE}\hlstd{)}\hlopt{$}\hlstd{class}
  \hlkwd{return}\hlstd{(}\hlkwd{confusionMatrix}\hlstd{(Ytest, ynew,} \hlkwc{negative}\hlstd{=negative))}
\hlstd{\}}

\hlcom{# set seed to get reproducible results}
\hlkwd{set.seed}\hlstd{(}\hlnum{1234}\hlstd{)}

\hlstd{cv.out} \hlkwb{<-} \hlkwd{crossval}\hlstd{(predfun.dda,}
                   \hlkwc{X}\hlstd{=featureMatrix[, top],}
                   \hlkwc{Y}\hlstd{=avgSpectra.info}\hlopt{$}\hlstd{health,}
                   \hlkwc{K}\hlstd{=}\hlnum{10}\hlstd{,} \hlkwc{B}\hlstd{=}\hlnum{20}\hlstd{,}
                   \hlkwc{negative}\hlstd{=}\hlstr{"control"}\hlstd{,}
                   \hlkwc{verbose}\hlstd{=}\hlnum{FALSE}\hlstd{)}
\hlkwd{diagnosticErrors}\hlstd{(cv.out}\hlopt{$}\hlstd{stat)}
\end{alltt}
\begin{verbatim}
      acc      sens      spec       ppv       npv       lor 
0.9500000 0.9000000 1.0000000 1.0000000 0.9090909       Inf 
\end{verbatim}
\end{kframe}
\end{knitrout}

As a result of the above analysis, we conclude that the identified peaks with
mass \ac{mz} $8937$ and $4467$ allow for the construction of a very low-dimensional
predictor
function that is highly effective in separating cancer and control group with both high accuracy as well as high sensitivity.

\section{Conclusion}

The large-scale acquisition of mass spectrometry data is becoming routine in 
many experimental settings.  In \Mq\ we have put together a robust \R\ pipeline for
preprocessing these data to allow
subsequent high-level statistical analysis.  All methods implemented in \Mq\ have
been selected both for computational efficiency and for biological validity.
In this chapter we have given an overview over the most commonly used procedures
of \Mq\ as well as demonstrated their application in detail.

A topic that has not been covered here is \acf{MSI}, which combines
spectral measurements with spatial information \citep{Cornett2007}.  
\Mq{} also enables some simple \ac{MSI} analysis, for
practical examples in  \R\
we refer to the homepage of \Mq\ at \url{http://strimmerlab.org/software/maldiquant/} 
as well as the associated web page
\url{https://github.com/sgibb/MALDIquantExamples/}.

\begingroup\renewcommand{\clearpage}{}

\renewcommand{\bibname}{\Large References}

\setlength{\bibsep}{0pt plus 0.3ex}

\bibliographystyle{apalike}
\bibliography{references}

\endgroup

\end{document}